\documentclass[letterpaper, 10 pt, conference]{ieeeconf}  

\IEEEoverridecommandlockouts                              
\usepackage[utf8]{inputenc}
\usepackage[tbtags]{amsmath}
\usepackage{bbm}
\usepackage[subnum]{cases}
\usepackage{amssymb, psfrag, epstopdf}

\usepackage{amsthm}
\usepackage{enumerate}
\usepackage{graphicx, subfigure}
\usepackage{color,cite,mathtools,booktabs}
\usepackage{cleveref}
\crefname{section}{§}{§§}
\Crefname{section}{§}{§§}
\usepackage{algorithmicx,algorithm,algpseudocode}

\newtheorem{thm}{Theorem}[section]
\newtheorem{lem}[thm]{Lemma}

\DeclareMathOperator*{\argmin}{arg\,min}
\newcommand{\squeezeup}{\vspace{-1.5 mm}}
\begin{document}
\title{Optimal Topology Design for Disturbance Minimization in Power Grids}
\author{Deepjyoti Deka$^{\dag}$, Harsha Nagarajan$^{\dag}$, Scott Backhaus$^{\dag}$ \thanks{Center for Nonlinear Studies, Los Alamos National Laboratory, NM, USA {\tt\small (deepjyoti,harsha,backhaus)@lanl.gov}. The work was supported by funding from the Advanced Grid Modeling Program in the Office of Electricity in U.S. Department of Energy.}}

\maketitle

\begin{abstract}
The transient response of power grids to external disturbances influences their stable operation. This paper studies the effect of topology in linear time-invariant dynamics of different power grids. For a variety of objective functions, a unified framework based on $H_2$ norm is presented to analyze the robustness to ambient fluctuations. Such objectives include loss reduction, weighted consensus of phase angle deviations, oscillations in nodal frequency, and other graphical metrics. The framework is then used to study the problem of optimal topology design for robust control goals of different grids. For radial grids, the problem is shown as equivalent to the hard ``optimum communication spanning tree" problem in graph theory and a combinatorial topology construction is presented with bounded approximation gap. Extended to loopy (meshed) grids, a greedy topology design algorithm is discussed. The performance of the topology design algorithms under multiple control objectives are presented on both loopy and radial test grids. Overall, this paper analyzes topology design algorithms on a broad class of control problems in power grid by exploring their combinatorial and graphical properties.
\end{abstract}

\section{Introduction}
\label{sec:introduction}
The power grid comprises of the network of generators, loads and transmission lines that enable the delivery of electricity. Operationally, the power grid is divided into several hierarchies. The long and medium interval voltages and powers at different generators and loads are decided in a settlement market. On the other hand, the dynamics of voltages and frequencies of grid nodes are represented by coupled swing equations that depend on the net power balance \cite{kundur}. If the net power is positive, the frequency increases and if it is negative, the frequency decreases. Automatic feedback loops and centralized signalling from the grid operator change the injected nodal powers to stabilize these dynamics. Such control is necessary as the  majority of devices in the power grid can safely operate only within a short guard band around their standard stable regime. In recent years, due to the proliferation of stochastic renewable resources and active loads, the fluctuations in operating frequency has increased and led to concerns about the stability of the grid \cite{ulbig2014impact}. Such problems have been compounded by the fact that new generation resources (including solar panels, wind resources) have lower rotating mass compared to traditional generators and thereby lower inertia to dynamics of the grid state variables.

Improving the dynamic performance of the modern grid has thus received greater attention in both academia and industry.
Efforts in this direction include analysis of techniques to control loads as an ancillary service \cite{seanmeyn}. Independent system operators (ISOs) have started payment structures and novel markets to incentivize new controllable resources \cite{ferc755, ercotwhitepaper, virtualinertia}. One of the techniques that can enable a grid to improve its dynamic performance using available resources is topology configuration. This refers to a pre-operation planned change in the grid topology. Note that topology reconfiguration of grids has been shown to help in grid resilience following natural calamities, address congestion issues with reduced wind curtailment as well as settlement in power markets  \cite{nagarajan2016optimal, dekaisgt,villumsen2013line}. In this paper, we are interested in quantifying its effect in improving system dynamics and utilizing it to design optimal topologies to achieve control goals. The topology of the grid (or any networked dynamic system) and its features (line impedances etc.) influence the flow of disturbances in the network and its stability. Past research within the power systems community has looked at specific goals (optimal loss reduction \cite{gayme}, improvement in feedback control \cite{john}, \cite{mallada}, consensus based network design \cite{SDPvoltage} and augmentation \cite{lygeros}) while analyzing topology reconfiguration using system theoretic tools.
There is also a line of work in topology design and augmentation for dynamic control using tools from semi-definite programming (SDP) \cite{arpita, SDPvoltage}.

In this paper, we study the effect of topology on dynamics of the power grid under general control goals that depend on the state variables (voltage phase angles, frequency). Such goals include loss reduction, fast damping of oscillations to common set point etc \cite{cps}. We use $\mathcal{H}_2$ norm of the system dynamics \cite{kundur} as a metric for performance that is commonly used to study the stability of linear dynamical systems. Under general cost functions, we show that the optimal topology reconfiguration can be formulated as a combinatorial optimization problem for \textit{weighted sum of effective inverse susceptances} in the graph. This holds for both radial (tree-structured) and loopy (with cycles) topologies. For radial topologies that relate to distribution grids, we show that the optimal configuration is given by the solution of the Optimum Communication Spanning Tree problem, a NP-hard problem in graph theory. We present a combinatorial algorithm for radial topology design with provable bounded approximation gap from optimality. Compared to existing works \cite{gayme, john, SDPvoltage} that use continuous valued relaxations, our formulation represents edge augmentation as a discrete option and is applicable for a wider range of operating conditions and parameter settings. For grids with loops, we use greedy augmentation schemes to design the topology of the grid from a constructed tree.

We test the performance of our algorithms (for radial and loopy networks) for physically relevant control goals on test power grid networks. To demonstrate their efficacy, we compare the performance with optimal topologies generated by brute-force enumeration over the set of permissible edges in the system. To summarize, we develop a theoretical framework for topology reconfiguration problems for power grids and develop topology design algorithms with bounded gap. We believe that hybrid approaches that combine inter-related areas (system theory, discrete optimization and graph theory) can enable significant improvements in grid design compared to individualistic methods.

The rest of the paper is organized as follows. Section \ref{sec:formulation} mathematically formulates our control problem under consideration. Section \ref{sec:properties} studies the properties of optimal topology design of grids. Sections \ref{sec:treenetwork} and \ref{sec:meshnetwork} provide a detailed analysis of greedy-based algorithms with approximation guarantees for tree and loopy networks, respectively. Section \ref{sec:nx} discusses a numerical study based on two test cases followed by conclusions and future directions in Section \ref{sec:conc}.

\section{Mathematical formulation}
\label{sec:formulation}
The power grid is represented mathematically as a connected graph ${\cal G }= \{{\cal V}, \cal{E}\}$,  where ${\cal V} = \{1,2,..N\}$ is the set of $N$ buses/nodes of the graph and ${\cal E} = \{e_{ij}\}$ is the set of undirected lines/edges. Let $y_{ij}=g_{ij}-\hat{j} b_{ij}$ denote the complex admittance of line $(ij)$ in the grid ($\hat{j}^2=-1$) with conductance $g_{ij} > 0$ and susceptance $b_{ij} > 0$. Each node $i$ is associated with a time varying complex voltage $V_i$ of magnitude $|V_i|$ and phase angle $\theta_i$. The frequency at node $i$ is denoted by $\omega_i$ where $\omega_i = \frac{\delta \theta_i}{\delta t}$. Under stable operating conditions, the frequencies at all nodes are maintained at a constant value $\omega^0 = 60\text{Hz}$ (in U.S.A.). The temporal dynamics of the grid is represented by the following swing equation \cite{kundur}:
\begin{align}
\footnotesize
M_i\dot{\omega}_i + D_i(\omega_i - \omega^0) = P^m_i-\sum_{(ij)\in {\cal E}}P_{ij}  +  u_i\label{swing}
\end{align}
Here, $M_i$ denotes the inertia of the rotating mass at node $i$, which primarily stems from inertia of generators. $D_i$ represents the damping at node $i$. $P^m_i$ represents the real power injection at the node. $P_{ij}$ is the real power flowing out of node $i$ through lines connected to its neighbors. Finally, $u_i$ is the external disturbance at node $i$. Using the DC power flow model \cite{kundur}, the power flow $P_{ij}$ on line $(ij)$ is given by
\begin{align}
\footnotesize
P_{ij}  = b_{ij}(\theta_i-\theta_j)\label{DCpowerflow}
\end{align}
In the absence of external disturbances, the system of equations has a stable operating point ($\omega_i =\omega^0, \dot{\omega}_i= 0$). As the swing equation is linear, we take the stable operating point as reference and express dynamics in terms of deviations from the reference. Abusing notation, from this point we use $\omega_i, \theta_i,P_{ij}$ to denote the deviation from their stable states. Using the linear power flow equations, the swing equation in vector form is as follows:
\begin{align}
[M]\dot{\omega} +  [D]\omega = -L_b\theta +u \label{swingmatrix}
\end{align}
Here $\omega,\theta, u$ are the $N{\times}1$ vectors. $[M]$ and $[D]$ are diagonal matrices representing the inertia and damping at nodes respectively. $L_b$ is the $N{\times}N$ susceptance weighted graph Laplacian of the grid $\cal G$ with the following structure:
\begin{align}
\footnotesize
    {\huge L_b}(i,j) = \begin{cases}-b_{ij} & \text{if $(ij) \in {\cal E}$}\\
\sum_{(ij)\in {\cal E}}b_{ij} & \text{if $j =i$}\\
0 & \text{otherwise} \end{cases} \label{treeinv}
\end{align}
Thus, the $i^{th}$ component of $L_b\theta$ gives the sum of outward power flows on all lines connected to node $i$.
Writing Eq.~(\ref{swingmatrix}) in standard $LTI$ system form \cite{kundur}, we have
\begin{align}
\footnotesize
   \begin{bmatrix} \dot{\theta} \\ \dot{\omega}  \end{bmatrix}   = \begin{bmatrix} 0 & \mathbb{I}\\ -M^{-1}L_b  & -M^{-1}D \end{bmatrix} \begin{bmatrix} \theta \\ \omega  \end{bmatrix} + \begin{bmatrix} 0\\ M^{-1}\end{bmatrix}u \label{swingmatrixfull}
\end{align}
where $\mathbb{I}$ is the identity matrix of size $N\times N$. Keeping our focus on topology design in the rest of the paper, we make the following assumption:

\textbf{Assumption 1}: At each node $i$, $M_i > 0$ (non-zero inertia), and $D_i = d$ (constant damping).

The assumption of non-zero inertia is made for convenience of presentation and our results hold for the case where certain nodes do not possess inertia. Further, using Kron-reduction \cite{florian} of graphs, a network can be reduced to only nodes with inertia. The second assumption is similar to the ones in previous work \cite{gayme, john}. In their absence, some equality results (noted in later sections) will be replaced by bounds. Note that, unlike prior work, no assumption is made on relative values of inertia at different nodes in the system.
\subsection{Control Goals}

\begin{figure}[!t]
    \centering
    \includegraphics[scale=0.85]{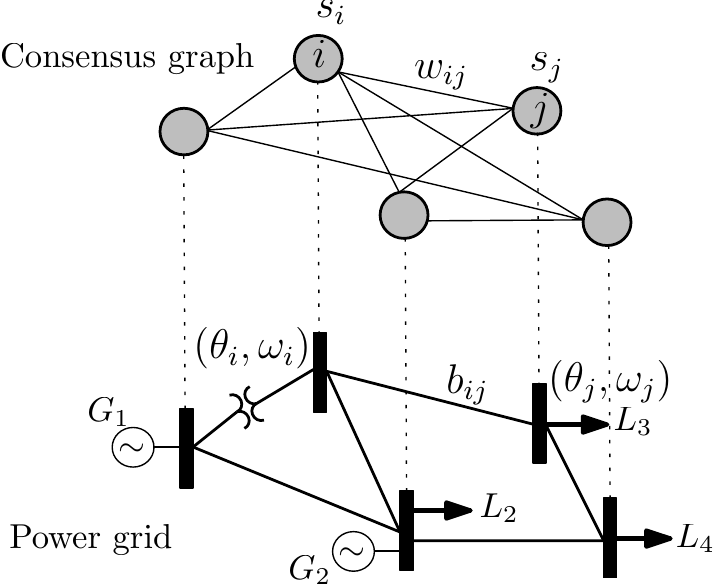}
    \caption{Illustration of power grid and its associated consensus graph.}
    \label{fig:concept}
    \vspace{-0.5cm}
\end{figure}

We now state a generalized set of control objectives that the observer/system operator is interested in optimizing, expressed as functions of the states ($\theta_i, \omega_i$) of the system at node $i$. Under ambient white noise $u$ and given $w_{ij}\geq 0, s_i \geq 0$, we are interested in minimizing the expected steady-state value of a non-negative function $f$ given by:
\begin{align}
   f := \sum_{\forall i \neq j} w_{ij}(\theta_i-\theta_j)^2 + \sum_{i \in \mathcal{V}}s_{i}\omega_i^2,
   \label{loss}
\end{align}

Note that $f$ depends on the differences of phase angles of nodes (not necessarily neighboring) and on the magnitude of frequency at each node. It follows immediately that $f = y^Ty$, the $\mathcal{L}_2$ norm of function $y$, where $y$  is given by:
\begin{align}
y =  \begin{bmatrix}L_w^{1/2}  ~&0 \\0 ~&S^{1/2}\end{bmatrix}  \begin{bmatrix}\theta\\ \omega\end{bmatrix} \label{output}
\end{align}
Here $L_w$ is the Laplacian matrix of graph ${\cal G}_w$ with edge weights given by $w_{ij}$. Note that ${\cal G}_w$ is not restricted to have the same topology as the power grid ${\cal G}$ (see Fig. \ref{fig:concept}). In fact, it can even be a complete graph. $S$ is a diagonal matrix with the $i^{th}$ diagonal entry given by $s_i$. As (weighted) Laplacian matrices are positive semi-definite, their matrix square roots ($L_w^{1/2}, S^{1/2}$) exist. Well-known functions for grid control can be derived using different choices of $w_{ij}$ ($L_w$) and $s_{i}$ ($S$) as listed:
\begin{enumerate}
    \item \textbf{Frequency Control}:  $w_{ij} = 0, s_{i} = 1$ ($S = \mathbb{I},~W = 0$).
    \item \textbf{Line Loss Reduction}: $s_{i} = 0, w_{ij} = g_{ij}\mathbf{1}((ij) \in {\cal E})$ ($S = 0,~ L_w  = L_g$, the conductance weighted Laplacian matrix for ${\cal G}$).
    \item \textbf{Consensus}: $w_{ij} = 1, s_{i} = 0$ ($S =0, ~L_w = \mathbb{I}- \textbf{1}\textbf{1}^T$).
\end{enumerate}

Further, we consider a modified \textbf{ranked consensus} control objective where each node $i$ in the grid is given a rank $r_i >0$ and the weight $w_{ij}$ for line $(ij)$ is given by the sum of the ranks, i.e., $w_{ij} = r_i+r_j$. This can help prioritize consensus between critical nodes over others. For example, if generators are ranked higher than loads, the addition of ranks will imply that $w_{ij}$ for a generator pair is greater than $w_{ij}$ for generator-load pair, and further greater than $w_{ij}$ for a load pair. Figure \ref{fig:concept} graphically illustrates the power grid and it's associated consensus graph for the grid control. For function $f$ given in Eq.~(\ref{output}), next we formulate the topology design problem.

\textbf{Topology Design Problem}: Given an input set of susceptance weighted edges ${\cal E}^{full}$ and budget $k \geq N-1$, an edge set ${\cal E}$ needs to be selected to optimize the following problem:
\begin{subequations}
\begin{align}
    \argmin_{{\cal E} \in {\cal E}^{full}} &\lim_{t\rightarrow \infty}\mathbb{E}\{y^T(t)y(t)\}\\
    s.t.~ & |{\cal E}| = k\label{cardinality}\\
          & [\theta,~\omega]^T \text{~satisfies Eq.~}(\ref{swingmatrixfull}), y \text{~satisfies Eq.~}(\ref{output})\\
          & {\cal E} \text{~is connected}
\end{align}
\label{problem}
\end{subequations}
Here Eq.~(\ref{cardinality}) reflects the budget on number of edges. If $k=N-1$, the problem is restricted to finding the optimal tree configuration. This is specifically important for radially operated distribution grids.

\textbf{Relation to Stability Analysis:} Note that the cost function in Eq.~(\ref{problem}) is exactly the squared \textbf{${\cal H}_2$} norm \cite{boyd1994linear} of a system where the dynamics are represented by Eq.~(\ref{swingmatrixfull}), while the observations are represented by Eq.~(\ref{output}).
\begin{align}
||H||^2_{{\cal H}_2} = \lim_{t\rightarrow \infty}\mathbb{E}\{y^T(t)y(t)\} \label{h2}
\end{align}
The popular ${\cal H}_2$ norm is used in the control community as a stability metric which has several interpretations. For unit impulse disturbances $u = \delta(t)$, ${\cal H}_2$ norm denotes the total variance of the output of the system to reach the steady state ($\int_{0}^{\infty} \mathbb{E}\{y^T(t)y(t)\}dt$). In our case, we consider the setting where the system is excited by persistent ambient noise $u$ of known variance.
In the next section, we use properties of ${\cal H}_2$ norm to determine a tractable form of the optimization objective for topology design.

\section{Optimal Topology Reconfiguration Problem}
\label{sec:properties}
The objective of the ISO, as discussed in the previous section, is to minimize the squared ${\cal H}_2$-norm of the dynamic swing equations. The ${\cal H}_2$ norm of a standard LTI system can be described by the following set of equations \cite{boyd1994linear}:
\vspace{-0.2cm}
\begin{align}
||H||^2_{{\cal H}_2} &= \text{Tr}(B^TQB)\label{observability1}\\
A^TQ+QA &= -C^TC \label{observability}
\end{align}
where ($A,B,C$) for our system are the matrices $\begin{bmatrix} 0 &I\\ -M^{-1}L_b  & -M^{-1}D \end{bmatrix}$, $\begin{bmatrix} 0\\ M^{-1}\end{bmatrix}$ and $\begin{bmatrix}L_w^{1/2}  ~&0 \\0 ~&S^{1/2}\end{bmatrix}$, respectively. Matrix $Q = \begin{bmatrix} Q_1 &Q_0\\ Q_0^*  & Q_2\end{bmatrix}$ is positive semi-definite and represents the observability Gramian of the system. As $M$ is a diagonal matrix in our case, the ${\cal H}_2$ norm reduces to
\begin{align}
||H||^2_{{\cal H}_2} & = \text{Tr}(M^{-1}Q_2M^{-1}) = \text{Tr}(M^{-2}Q_2)\label{H_2eq}
\end{align}
Eq.~(\ref{observability}) can be expanded to:
\begin{subequations}
\footnotesize
\begin{align}
&\begin{bmatrix} 0 &-L_bM^{-1}\\\mathbb{I}   & -DM^{-1} \end{bmatrix}\begin{bmatrix} Q_1 &Q_0\\ Q_0^*  & Q_2\end{bmatrix} +\nonumber\\ &\begin{bmatrix} Q_1 &Q_0\\ Q_0^*  & Q_2\end{bmatrix}\begin{bmatrix} 0 &\mathbb{I}\\ -M^{-1}L_b  & -M^{-1}D \end{bmatrix} =- \begin{bmatrix}L_w  ~&0 \\0 ~&S\end{bmatrix} \label{equalities}
\end{align}
\end{subequations}
This gives us four equality relations, one for each submatrix. Since trace is invariant under cyclic permutations, the first equality reduces to

\begin{subequations}
\footnotesize
\begin{align}
&L_b M^{-1}Q_0^*+ Q_0M^{-1}L_b = L_w\\
\Rightarrow~&2\text{Tr}(Q_0M^{-1})= \text{Tr}(L_wL_b^+)\label{first_equality}
\end{align}
\end{subequations}
Multiplying $M^{-1}$ on both sides of the fourth equality from Eq.~(\ref{equalities}) and using Eq.~(\ref{first_equality}), we have

\begin{subequations}
{\footnotesize
\begin{align}
&(DM^{-1}Q_2+ Q_2M^{-1}D)M^{-1}-(Q_0+ Q_0^*)M^{-1} = SM^{-1}\nonumber\\
&\Rightarrow~ \text{Tr}(M^{-2}Q_2) = (\text{Tr}(L_wL_b^+) + \text{Tr}(SM^{-1}))/2d \label{combine}
\end{align}}
\end{subequations}
The closed form of the squared ${\cal H}_2$ norm of the system is summarized in the following lemma.
\begin{lem}\label{H_2lemma}
The squared ${\cal H}_2$ norm of the LTI dynamical system given by Eqs.~(\ref{swingmatrixfull},\ref{output}) is given by $(\text{Tr}(L_wL_b^+) + \text{Tr}(SM^{-1}))/2d$.
\end{lem}
Without Assumption $1$ (constant damping), Eq.~(\ref{combine}) will provide a bound on the squared ${\cal H}_2$ norm based on the maximum and minimum values of nodal damping. A similar albeit restricted formulation is derived in \cite{gayme} where Laplacian matrices $L_w$ and $L_b$ are defined over the same graph, assuming equal nodal inertias.

Note that in Lemma \ref{H_2lemma}, the effect of $S$ (weights associated with frequency deviations) on the ${\cal H}_2$ norm is separable from that of the grid topology (or $L_b$). Thus, the search for the optimal grid topology can be limited to the first term of Lemma \ref{H_2lemma}. This provides the following reformulation of Problem \ref{problem}:
\begin{subequations}
\begin{align}
    \argmin_{{\cal E} \in {\cal E}^{full}} ~&\text{Tr}(L_wL_b^+)\\
     \text{s.t.}~ & |{\cal E}| = k, \  rank(L_b) = N-1 \label{rank}
\end{align}
\label{problem_1}
\end{subequations}
The $N-1$ rank constraint for $L_b$ ensures that the graph constructed using set $\cal E$ has one connected component. Note that brute force schemes to determine the optimal topology quickly become intractable due to the exponential number of candidate feasible graphs that can be constructed.

\subsection{Pseudo-inverse and graph distances}
We now use properties of Laplacian pseudo-inverse to describe Problem \ref{problem_1} in terms of effective graph distances. This will enable us to relate Problem \ref{problem_1} to studied problems in graph theory. By definition, we define $L_b^+(i,i)+ L_b^+(j,j) - 2L_b^+(i,j)$ as \textbf{effective inverse susceptance ($b^{-1}_{eff}(i,j)$)} between nodes $i$ and $j$ in the graph. Effective inverse susceptance (similar to effective resistance) in the DC power flow model represents the ratio between phase angle difference between nodes $i$ and $j$ when one unit of active power is inserted at $i$ and taken out at $j$. We can expand the cost function in Problem \ref{problem_1} as follows:

\begin{subequations}
{\footnotesize
\begin{align}
\text{Tr}(L_wL_b^+) &= \sum_{\forall i \neq j}w_{ij}(L_b^+(i,i)+ L_b^+(j,j) - 2L_b^+(i,j))\label{graph_eq1}\\
&= \sum_{\forall i \neq j}w_{ij}b^{-1}_{eff}(i,j) \label{graph_eq}
\end{align}}
\end{subequations}
The optimal topology design problem \ref{problem_1} can now be interpreted as:
\begin{lem}\label{graph_distance}
The cost function in Problem \ref{problem_1} is equivalent to \textit{minimizing the weighted sum of effective inverse susceptances between all pairs of nodes in the graph}.
\end{lem}
The expression in Eq.~(\ref{graph_eq}) can also be listed in terms of the full-rank \textit{reduced} Laplacian matrices $\hat{L}_w, \hat{L}_b$ of size $(N-1)\times (N-1)$. In the next two sections, we analyze two variants of the optimal topology design  problem: one for radial networks and the other for meshed networks, and discuss their computational hardness and solution schemes.

\section{Optimal Tree Construction Problem}
\label{sec:treenetwork}
We analyze the case where the number of graph edges is $N-1$, i.e., the constructed graph is a tree. Note that constraining the network to be a tree is a common objective in distribution grids, which are historically operated in a radial configuration.

\textbf{Reformulation}: In this case, the optimal $L_b$ to be constructed in Problem \ref{problem_1} and Eq.~(\ref{graph_eq}) corresponds to edge set ${\cal E}^{\cal T}$ for some tree ${\cal T}$. Using properties of pseudo-inverse in trees \cite{dekatcns}, it can shown that the effective inverse susceptance $b^{-1}_{eff}(i,j)$ for two nodes $i$ and $j$ in a radial graph has a simple expression - it is equal to the graph distance $d^{\cal T}_{ij}$ between nodes $i$ and $j$ if each edge $e$ in ${\cal T}$ is given distance $d_e =1/b_e$. 
We thus have the following reformulation of the topology design Problem \ref{problem_1} for radial networks:
\begin{subequations}
\begin{align}
     \argmin_{{\cal E}^{\cal T} \in {\cal E}^{full}} ~&\sum_{\forall i,j}w_{ij}d^{\cal T}(i,j) \\
     \text{s.t}.~ & {\cal E}^{\cal T} \text{~forms a tree}
\end{align}
\label{problem_tree}
\end{subequations}
Using this reformulation, we are able to connect Problem \ref{problem_tree} to a NP-hard problem in graph theory.
\begin{thm}\label{comm_tree}
The optimal radial grid that minimizes the ${\cal H}_2$-norm based cost function in Eq.~\ref{loss} is given by the solution to the \textbf{Optimal Communication Spanning Tree} Problem \cite{hu1974optimum} in graph theory.
\end{thm}
The Optimum Communication Spanning Tree problem in \cite{hu1974optimum} determines the spanning tree that minimizes the sum of communication between all nodes, where the cost of communication between a node pair $i$ and $j$ is given by a constant multiplied by the sum of edge distances on the unique path connecting them. The Optimum Communication Spanning Tree problem can be shown to be NP-hard in general.

\textbf{Algorithm for Optimal Tree:} We now discuss a scheme for topology design for Problem \ref{problem_tree}. We extend results in \cite{Yyu1} to prove the approximation gap of our scheme. In particular, we consider \textit{rooted shortest-path trees} in $\tilde{{\cal G}}$, the graph formed by all candidate edges in ${\cal E}^{full}$. A shortest path tree ${\cal T}$ rooted at node $k$ is a spanning tree in ${\cal E}^{full}$ such that for each node $i$, distance $d^{\cal T}(i,k)$ is equal to the distance $d^{\tilde{{\cal G}}}(i,k)$ on the shortest path from $i$ to $k$ in $\tilde{{\cal G}}$. The shortest-path tree with minimum total distance $\sum_{i} d^{\cal T}(i,m)$ is called the \textbf{minimum shortest path tree} and its root $m$ is called the \textbf{`median'}. The following result (extends Lemmas $3$ and $5$ in \cite{Yyu1}) upper bounds the performance of the minimum shortest path tree for the objective in Problem \ref{problem_tree}.

\begin{lem}\label{upper}
The minimum shortest path tree $\cal T$ rooted at median $m$ of graph satisfies $\tilde{{\cal G}}$, $\sum_{\forall i,j}w_{ij}d^{\cal T}(i,j) \leq \sum_i d^{\tilde{{\cal G}}}(i,m) \max_i(\sum_jw_{ij})$.
\end{lem}
The proof is omitted due to space constraints. Further it can be shown that the following holds
{\footnotesize
\begin{align}
    &\sum_{\forall i,j}w_{ij}d^{{\cal T}^*}(i,j) \geq \sum_{i} d^{\tilde{{\cal G}}}(i,m) \min_i \left(\smashoperator[r]{\sum_{j \in S,|S|=N/2}}w_{ij}\right)\label{lower}
\end{align}}
The derivation of the above statement is postponed to the general version for space constraints. Combining Lemma \ref{upper} and inequality (\ref{lower}),we have
\begin{lem}\label{gap}
Let ${\cal T}^*$ be the optimal tree solution to Problem \ref{problem_tree}. The minimum shortest path tree $\cal T$ rooted at median $m$ of graph $\tilde{{\cal G}}$ has approximation gap from optimality given by:
$$  \frac{\sum_{\forall i,j}w_{ij}d^{\cal T}(i,j)}{\sum_{\forall i,j}w_{ij}d^{{\cal T}^*}(i,j)}  \leq \frac{\max_i(\sum_j w_{ij})}{ \min_i(\smashoperator[lr]{\sum_{j \in S,|S|=N/2}}w_{ij})}$$
\end{lem}

For two specific cases discussed in the Introduction (\textbf{consensus} and \textbf{ranked consensus}), we can show that \textbf{approximation gap} given by Lemma \ref{gap} is bounded by $2$. The tree construction steps are listed  in Algorithm $1$. We select node $k$ that minimizes the objective of Problem \ref{problem_tree} as root as it has performance at par or better than the median rooted tree.
\begin{algorithm}
\footnotesize
\caption{Radial Grid Construction for Problem \ref{problem_1}}
\textbf{Input:} Set of permissible susceptance ($b$) weighted edges ${\cal E}^{full}$ over nodes in set ${\cal V}$, weights $w_{ij}$ for node pairs $(i, j)$\\
\textbf{Output:} Tree ${\cal T}$
\begin{algorithmic}[1]
\State Get graph $\hat{\cal G}$ of all edges in ${\cal E}^{full}$. Compute weighted Laplacian matrix $L_w$ using non-zero $w_{ij}$ as weighted edges, $temp \gets \infty, k \gets 1$.
\ForAll{$i \in \hat{\cal G}$}
    \State Find susceptance weighted Laplacian $L_b$ of shortest path tree rooted at node $i$ in $\hat{\cal G}$.
    \If{$\text{Tr}(L_wL_b^+) < temp$}
        \State $k \gets i$
    \EndIf
\EndFor
\State ${\cal T} \gets$ shortest path tree rooted at $k$
\end{algorithmic}
\end{algorithm}

\textbf{Computational Complexity}: In terms of the number of nodes $N$, the complexity of Algorithm $1$ is $O(N^3)$ as the for loop iterates over $N$ nodes, while each rooted spanning tree computation takes $O(N^2)$ steps \cite{Yyu1}, when shortest path lengths to all nodes are known.
\section{Optimal Meshed Network Construction Problem}
\label{sec:meshnetwork}
In this section, we look at Problem \ref{problem_1} where the designed edge set $\cal E$ has cardinality $k > N-1$ and hence creates a meshed grid. Note that unlike a tree grid, the effective inverse susceptance $b^{-1}_{eff}(i,j)$ (see Eq.(~\ref{graph_eq})) for a meshed network is not a simple linear function of the line susceptances. To bring tractability, we design the meshed network as a two step process: (a) construct a tree ${\cal T}$ with $N-1$ edges, and then  (b) add $k-(N-1)$ edges to tree $\cal T$. The first step can be designed using Algorithm $1$ in Section  \ref{sec:treenetwork}. The second step involves adding $k-(N-1)$ edges to the tree, such that Laplacian matrix $L_b$ for edges minimizes $\text{Tr}(L_wL_b^+)$. We use a greedy algorithm to determine additional edges. \cite{submodularcorrection} shows that the optimization problem here is not supermodular \cite{submodulargreedy} in general and hence strong theoretical guarantees may not be permissible. In next section, we show the effectiveness of the greedy approach through simulations.

\textbf{Algorithm for Meshed Network:} Algorithm $2$ builds a meshed network of $k>N-1$ edges from a set ${\cal E}^{full}$. First we use Algorithm $1$ to construct tree  $\cal T$ and then greedily add $k-(N-1)$ edges to minimize the objective $\text{Tr}(L_wL_b^+)$.

\begin{algorithm}
\footnotesize
\caption{Meshed Grid Construction for Problem \ref{problem_1}}
\textbf{Input:} Set of permissible susceptance ($b$) weighted edges ${\cal E}^{full}$ over nodes in set ${\cal V}$, weights $w_{ij}$ for node pairs $(i, j)$, Number of graph edges $k$\\
\textbf{Output:} Graph ${\cal G}$
\begin{algorithmic}[1]
\State Use Algorithm $1$ to generate tree ${\cal T}$ with edges ${\cal E}^{\cal T}$
\State ${\cal E}^{\cal G} \gets {\cal E}^{\cal T}$
\For{$n \in \{0,k - (N-1)\}$}
 \State Find edge $e \in {\cal E}^{full} - {\cal E}^{\cal G}$ that minimizes $\text{Tr}(L_wL_b^+)$ where $L_b$ is susceptance weighted Laplacian formed by ${\cal E}^{\cal G} \cup \{e\}$.
 \State ${\cal E}^{\cal G} \gets {\cal E}^{\cal G} \cup \{e\}$
\EndFor
\State Form $\cal G$ using edges in ${\cal E}^{\cal G}$
\end{algorithmic}
\end{algorithm}
\textbf{Computational Complexity}: Using Algorithm $1$ to form tree $\cal T$, the overall complexity of Algorithm $2$ is given by $0(N^3) + 0(kN^2)$.

\section{Numerical Simulations}
\label{sec:nx}
\begin{figure}[h]
    \centering
    \includegraphics[scale=0.57]{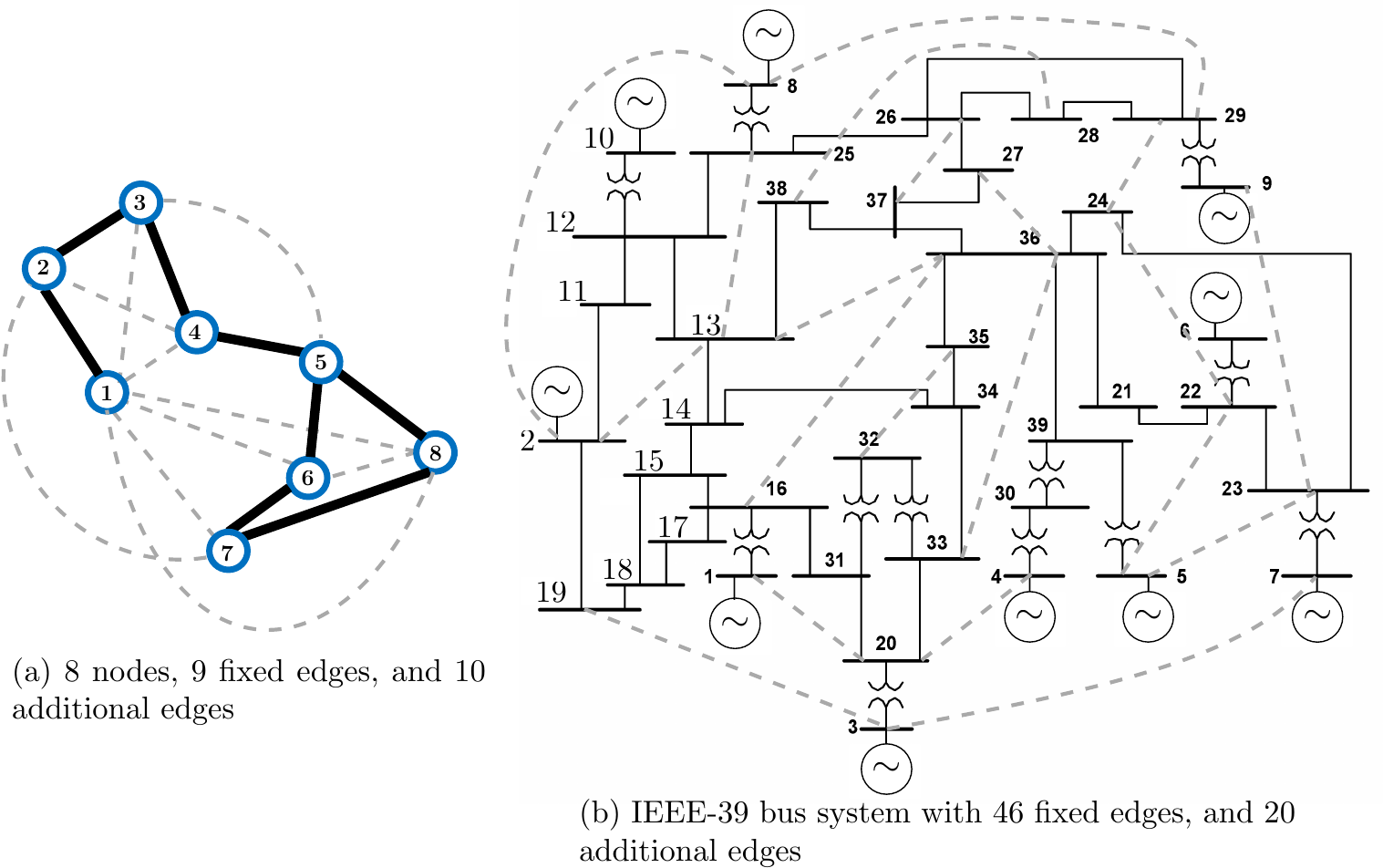}
    \caption{Test power grid networks. Additional edges (dotted) are randomly assigned to the base network.}
    \label{fig:cases}
    \vspace{-0.35cm}
\end{figure}
In this section, we demonstrate the performance of Algorithms $1$ and $2$ in designing radial and meshed networks to minimize two different cost functions for Problem \ref{problem_1}. The cost functions used are : (a) consensus ($w_{ij} = 1 \forall i \neq j$) and (b) ranked consensus ($w_{ij} = r_i+r_j$). We first test our algorithms on a $8$ node sub-graph of the IEEE $39$-bus test case \cite{39bus}. We use a permissible edge set ${\cal E}^{full}$ of cardinality $18$ (see Fig.~(\ref{fig:cases} (a)). Out of the $18$ edges, $8$ are derived from the original IEEE case, while the rest are randomly assigned. We consider the consensus cost and design networks with $7, 8, 9$ and $10$ edges selected from ${\cal E}^{full}$. To grade a constructed network, we determine the relative difference between its squared ${\cal H}_2$ norm  and that of the optimal network of same edge cardinality determined by enumeration/brute-force. The results are summarized in Table \ref{table_1}.

\begin{table}[ht]
\footnotesize
\caption{Relative Performance Gap of designed grids for consensus cost in $8$ node network. GA: Greedy Augmentation, OA: Optimal Augmentation.}
\squeezeup
\begin{tabular}{lcccc}
\toprule
\# of & Rooted Tree & Rooted Tree & Optimal Tree & Optimal Tree\\
edges & + GA(\%) & + OA (\%) &+ GA (\%) &+ OA (\%) \\
\cmidrule{1-5}
$7$ (tree)  &   $0.0189$  & $0.0189$  & $0$  &  $0$     \\
$8$  &  $6.8742$  &$  6.8742$  &$ 6.8743$  &$ 6.8743$   \\
$9$ &  $19.3806$  &$ 19.3806$  &$19.3790$  &$ 19.3790$    \\
$10$ &  $25.4962$ &$  25.4956$ &$ 25.4955$ &$  25.4955$   \\
\bottomrule
\end{tabular}
\label{table_1}
\vspace{-0.6cm}
\end{table}
\begin{figure}[h]
    \centering
    \includegraphics[scale=0.34]{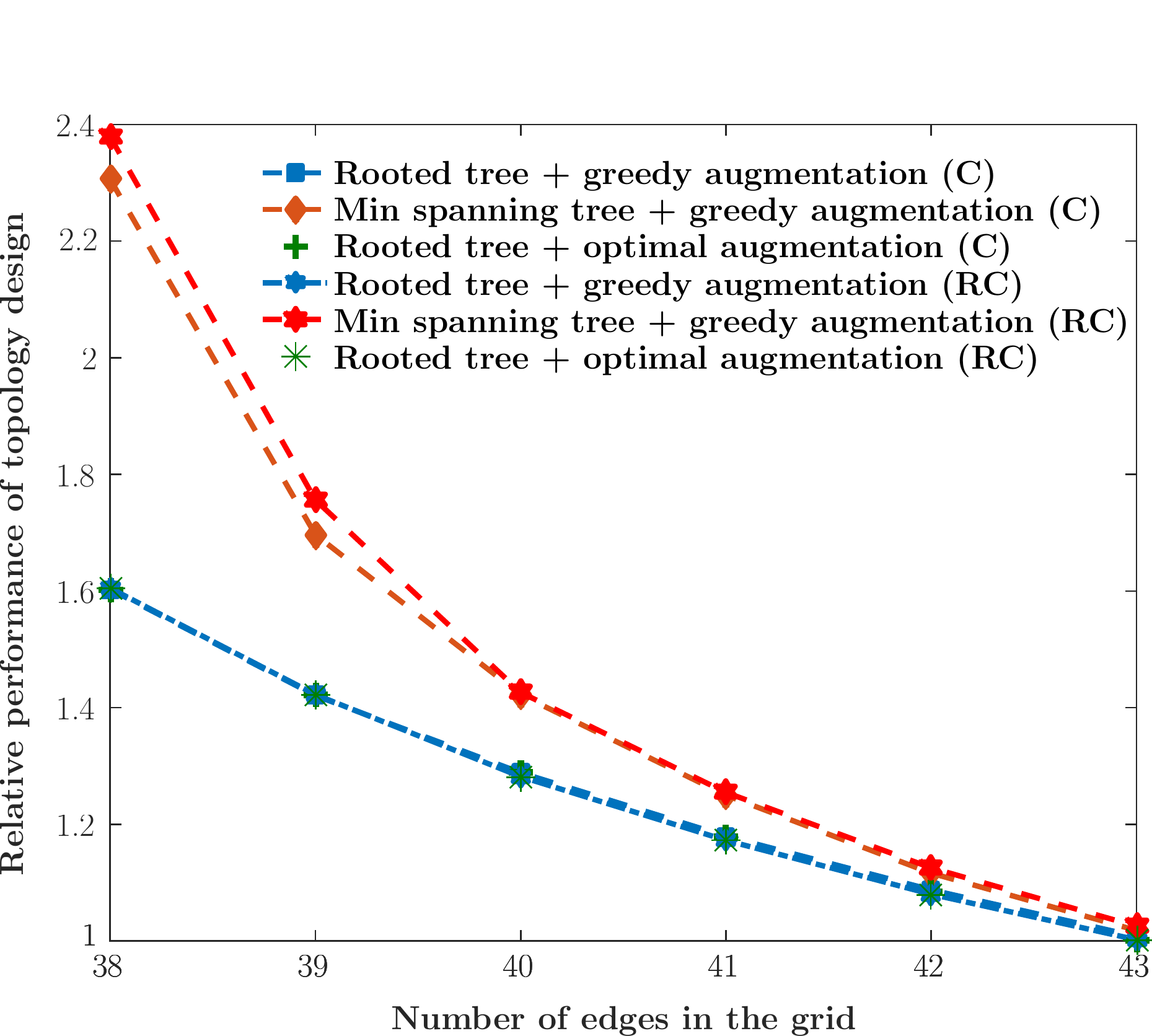}
    \vspace{-0.25cm}
    \caption{Performance of grid design for different edge cardinalities in the 39-bus system with consensus (C) and ranked consensus (RC) cost functions. Performance refers to ${\cal H}_2$-norm-based cost relative to the cost of optimal augmented network of $43$ edges. Lower relative cost implies higher performance.}
    \label{fig:39busresult}
    \vspace{-0.3cm}
\end{figure}

In Table \ref{table_1}, the second column reflects the performance of best rooted tree designed with Algorithm $1$ with edge augmented greedily using Algorithm $2$. On the other hand, the third column reflects the performance achieved by brute-force addition of edges to the rooted tree given by Algorithm $1$. Finally, in fourth and fifth columns, we consider the optimal tree generated by enumeration. The edge addition to the optimal tree is conducted via Algorithm $2$ in column $4$ and by brute force search in column $5$. The first entry in second column shows the relative performance difference for tree designed by Algorithm $1$ is less than $.02\%$. Similarly, comparing entries in column $2$ with column $3$, and column $4$ with $5$, we notice that the performance of greedy augmentation is comparable to that of brute force search. Finally, when the number of edges in the system is increased, the performance gap with the global optimal solution increases. Next, we consider the full IEEE $39$-bus test-case (see Fig.~\ref{fig:cases} (b)) with ${\cal E}^{full}$ composed of $66$ edges. Fig.~(\ref{fig:39busresult}) demonstrates the performance in designing networks of different cardinalities ($38$ (tree), $39-43$ (mesh)) for both consensus (C) and ranked consensus (RC) based cost functions. For both cost functions, we use Algorithm $1$ to first construct a rooted-tree and then select extra edges either greedily (Algorithm $2$) or by a brute-force search. Fig.~(\ref{fig:39busresult}) shows that the performance of greedy augmentation is comparable to that of brute force augmentation for all considered edge cardinalities under both cost functions. Further we present the performance of greedy addition of edges to the min spanning tree constructed from set ${\cal E}^{full}$ and show that it is outshone by topologies designed by our algorithms.

\section{Conclusions and Future Work}
\label{sec:conc}
This paper presents a general framework to study the effect of network topology on the dynamics of power grid. Combining ideas from control theory, algebraic graph theory and discrete optimization, we categorize the problem of optimal topology design to optimize a broad class of critical control objectives in the grid. We show the NP-hardness of the hardness of the topology design problem for both radial and meshed networks and discuss efficient algorithms and their computational complexity. For radial grid, we present a rooted spanning tree based topology design algorithm and demonstrate its approximation gap. For the meshed network, we discuss the application of greedy design algorithms to augment the topology beginning with a tree. The good performance and optimality gap of our topology design algorithms is presented on two test networks.

Future research in this area will include application of our algorithms to real-data sets and development of optimization algorithms for solving nonlinear mixed-integer (MI) SDPs based on the recent developments in the literature \cite{nagarajan2015maximizing,nagarajan2016tightening}.

\bibliographystyle{IEEETran}
\bibliography{references}
\end{document}